\documentclass[11pt,letterpaper]{article}
\usepackage[letterpaper,top=0.91in,bottom=0.92in,left=1.02in,right=1.02in]{geometry}
\usepackage[T1]{fontenc}
\usepackage{amsmath,amssymb,amsthm,mathtools}
\usepackage{newtxtext,newtxmath}
\usepackage{microtype}
\usepackage{bm}
\usepackage{graphicx}
\usepackage{cite}
\usepackage{booktabs}
\usepackage{xcolor}
\usepackage{tikz}
\usepackage{titlesec}
\usepackage{subcaption}
\usepackage{enumitem}
\usepackage{float}
\usepackage[section]{placeins}
\usepackage[hidelinks]{hyperref}
\hypersetup{pdftitle={Effective DoF Collapse in Sparse Near-Field Beamforming: A Block-Rank-1 Mechanism},
 pdfauthor={Changhao He, Xiaojuan Zhang, Francois Chin Po Shin},
 pdfsubject={Accepted manuscript, IEEE Wireless Communications Letters}}
\graphicspath{{./}{../wcl_sim/v4/}}
\usetikzlibrary{calc,arrows.meta,positioning,decorations.pathreplacing,patterns,shapes.geometric}

\titleformat{\section}{\large\bfseries}{\thesection.}{0.75em}{}
\titleformat{\subsection}{\normalsize\bfseries}{\thesubsection}{0.75em}{}
\titlespacing*{\section}{0pt}{3.1ex plus 1.2ex minus 0.5ex}{1.3ex plus 0.4ex}
\titlespacing*{\subsection}{0pt}{2.2ex plus 0.8ex minus 0.5ex}{0.9ex plus 0.3ex}
\newtheoremstyle{wclplain}{9pt plus 2pt minus 1pt}{9pt plus 2pt minus 1pt}
 {\itshape}{} {\bfseries} {.} {0.65em} {}
\newtheoremstyle{wclremark}{9pt plus 2pt minus 1pt}{9pt plus 2pt minus 1pt}
 {\normalfont}{} {\bfseries} {.} {0.65em} {}
\theoremstyle{wclplain}
\newcommand{\bp}{\bm{p}}
\newcommand{\bw}{\bm{w}}
\newcommand{\ba}{\bm{a}}
\newcommand{\bA}{\bm{A}}
\newcommand{\bP}{\bm{P}}
\newcommand{\bq}{\bm{q}}
\newcommand{\bI}{\bm{I}}
\newcommand{\bU}{\bm{U}}
\newcommand{\bu}{\bm{u}}

\newcommand{\bR}{\bm{R}}
\newcommand{\bC}{\bm{C}}
\newcommand{\bX}{\bm{X}}

\newcommand{\bc}{\bm{c}}
\newcommand{\br}{\bm{r}}

\newcommand{\Rg}{\bm{R}_g}
\newcommand{\Rd}{\bm{R}_{\rm dark}}
\newcommand{\Pill}{\bP_{\rm ill}}
\newcommand{\Pdark}{\bP_{\rm dark}}
\newcommand{\Cset}{\mathbb{C}}
\newcommand{\Rset}{\mathbb{R}}
\newcommand{\Gset}{\mathcal{G}}
\newcommand{\Pset}{\mathcal{P}}
\newcommand{\Mtot}{M_{\rm tot}}
\newcommand{\reff}{r_{\rm eff}}
\newcommand{\rprolate}{r_{\rm prolate}}
\newcommand{\Range}{\mathrm{Range}}

\newcommand{\tr}{\operatorname{tr}}
\newcommand{\avg}[1]{\langle #1 \rangle_{\Gset}}
\newtheorem{theorem}{Theorem}
\newtheorem{corollary}{Corollary}
\newtheorem{lemma}{Lemma}
\newtheorem{proposition}{Proposition}
\theoremstyle{wclremark}
\newtheorem{remark}{Remark}
\theoremstyle{wclplain}

\newcommand{\rhoavg}{\rho^\star_{\rm avg}}
\newcommand{\rhoeps}{\rho_\epsilon}
\newcommand{\etatot}{\eta_{\rm tot}}
\title{\vspace{-1.7em}
{\normalfont\scriptsize\itshape
This paper has been accepted for publication in IEEE Wireless Communications Letters (WCL). Copyright \copyright{} 2026 IEEE; copyright transferred to IEEE.\par}
\vspace{1.35em}
{\LARGE\bfseries Effective DoF Collapse in Sparse Near-Field Beamforming:\\[3pt]
A Block-Rank-1 Mechanism}}
\author{\normalsize Changhao He$^{1}$, Xiaojuan Zhang$^{2}$ (Senior Member, IEEE),\\
\normalsize Francois Chin Po Shin$^{2}$\\[7pt]
\small $^{1}$King Abdullah University of Science and Technology (KAUST), Saudi Arabia\\
\small $^{2}$A*STAR Institute of Advanced Intelligence and Computing (A*STAR IAIC), Singapore\\[3pt]
\small \texttt{changhao.he@kaust.edu.sa}; \texttt{\{xzhang,Francois\_Chin\}@a-star.edu.sg}\\[2pt]
\small Corresponding author: Xiaojuan Zhang}
\date{}

\begin{document}
\maketitle
\vspace{-1.15em}
\begin{abstract}
Sparse near-field arrays phase-cohere $M$ widely separated sub-arrays
of $N$ elements and underpin distributed MIMO, satellite formations,
and integrated sensing and communication. We show that their
interference suppression over a protected region is set by coherence
within a single sub-array rather than by the span of the formation:
the effective rank of the region covariance can collapse from $MN$
to $M$. Two independent arguments establish this. For canonical line
and disk regions each diagonal covariance block is exactly a
classical concentration operator,
which supplies sharp per-block rank thresholds and makes fewer than
one spatial degree of freedom per sub-array a conservative design
rule; an orthogonal decomposition with eigenvalue interlacing gives
a nonasymptotic sufficient condition that assumes nothing about the
coupling between platforms. The consequences are covariance-side and
beamformer-independent: the unloaded Capon functional upper-bounds
worst-case signal-to-interference ratio, while its regularized
counterpart scales very differently for angularly separated and for
in-cone targets. An eight-satellite LEO study validates the theory,
including realized worst-case SIR.
\end{abstract}

\noindent\textbf{Index Terms---}Distributed MIMO, sparse near-field arrays,
beamforming, covariance rank, effective DoF, prolate spheroidal functions.

\par\smallskip
{\footnotesize\noindent\textit{Funding:} This work was supported by the Agency
for Science, Technology and Research (A*STAR), Singapore, under the
Semiconductor RIE Flagship Programme (Integrated Circuit Design \& Systems),
grant No. H26-MSE1535.\par}
\vspace{0.35em}

\section{Introduction}
\label{sec:intro}

Distributed apertures that phase-cohere $M$ widely
separated sub-arrays are central to cell-free
MIMO, low-Earth-orbit (LEO) formation
beamforming~\cite{ref:satform}, modular extremely-large-scale
arrays~\cite{ref:liModularFocus2024}, and integrated sensing and
communication. Their defining feature is a \emph{mixed} propagation regime. For
the LEO system of Section~\ref{sec:numerics} the collective
Fraunhofer distance is
$2D^2/\lambda\approx1.2\times10^{9}$~km, so any user lies deep in
the collective radiative \emph{near field} (spherical wavefronts
across platforms), while each sub-array satisfies
$2(2R_a)^2/\lambda\approx101$~m~$\ll$~range and operates in its own
\emph{far field}: the ``sub-array far field, whole-array near
field'' regime~\cite{ref:liModularFocus2024}. For filled
apertures, classical theory ties suppression to aperture: enlarging
$D$ sharpens the Rayleigh resolution and the resolving power of
optimum inverse-covariance processors~\cite{ref:cox1973}, and
spreading antennas out helps despite spatial
aliasing~\cite{ref:bjornsonReality2019}. Whether this persists for a
sparse \emph{union} of $M$ small sub-apertures is the question here.

We study protected-region beamforming: maximize the gain toward target $\ba_t=\ba(\bq^\star)$ while suppressing the radiated field
over a protection region $\Gset\subset\Rset^3$ under a power
budget~\cite{ref:satform},

\begin{equation}\label{eq:socp}
\begin{aligned}
  \max_{\bw,\,t_g\ge 0}\; &\mathrm{Re}\{\ba_t^H\bw\}-\beta t_g \\
  \text{s.t.}\; &|\ba(\bq)^H\bw|\le t_g,\;\forall \bq\in\Gset, \quad
   \|\bw\|_2^2\le P_{\rm tot},
\end{aligned}
\end{equation}
with $\bw\in\Cset^{\Mtot}$, $\Mtot=MN$. The analysis is organized
around the region covariance $\Rg$ of~\eqref{eq:Rg-def} and the
Capon functional
$\rhoavg=\sup_{\bw\ne\bm0}|\ba_t^H\bw|^2/(\bw^H\Rg\bw)$: the exact optimum
of the \emph{average}-leakage criterion and, by
Proposition~\ref{prop:dual}, an upper bound on the
\emph{worst-case} SIR of every beamformer, including the optimum
of~\eqref{eq:socp}. The spectrum of $\Rg$ sets the DoF available
for suppression.

How many significant eigenvalues should $\Rg$ have? For a filled
aperture the count is the spatial Shannon number (aperture in
wavelengths times angular extent~\cite{ref:landauWidom1980}).
Applied to the \emph{convex
hull} of a sparse formation it suggests
$\rprolate\approx D^2\Omega_\Gset/\lambda^2\sim10^{10}$ modes at
the LEO anchor; for a union of separated compact sub-apertures
concentration theory scales instead with the \emph{occupied}
aperture measure, about $M$ times the per-sub-array count. What concentration theory
does not by itself determine is how the finite-$c$ per-block
eigenstructure combines with the inter-platform \emph{spherical}
phases (outside the Fourier setting
of~\cite{ref:landauWidom1980}), nor the resulting suppression
limits of the coupled sparse array. This letter supplies both.
Theorem~\ref{thm:rank1} gives an exact orthogonal decomposition of
$\Rg$ and a Cauchy-interlacing bound
$\lambda_{M+1}(\Rg)\le MN\sin^2\eta$, $\eta$ the intra-sub-array
phase excursion~\eqref{eq:A}, hence
$\reff(\Rg)\le M$ whenever $\tan^2\eta<1/(2M)$: no asymptotics, no
assumption on the coupling. Lemma~\ref{lem:coh} and Proposition~\ref{prop:slepian} identify
each diagonal block as a concentration operator of parameter
$c\le\eta$ \emph{exactly}: $\eta=\pi/2$ is
the coherence limit, a conservative design rule, while the
half-power block rank exceeds one only beyond $c^\star{=}2.28$ (line
region) or $c^\star_{\square}{=}2.52$ (square panel, disk region).
Corollaries~\ref{cor:invq}--\ref{cor:agnostic} and
Proposition~\ref{prop:dual} then give the beamforming ceilings,
validated on an eight-satellite LEO formation including
worst-case SIR.

\section{System Model}
\label{sec:model}

\subsection{Geometry, Steering Vectors, and Covariance}
Fig.~\ref{fig:system} shows the geometry. $M$ platforms at
$\{\bp_m\}_{m=1}^M\subset\Rset^3$ span
$D\triangleq\max_{m\ne n}\|\bp_m-\bp_n\|$. Sub-array $m$ has $N$
elements at $\bp_{m,k}=\bp_m+\bm{\xi}_k$, of edge length $D_a$ and
\emph{circumradius} $R_a\triangleq\max_k\|\bm{\xi}_k\|\ll D$, so
$R_a=D_a/2$ for a linear sub-array and $D_a/\sqrt2$ for a square
panel; in the numerics each is a $\sqrt N\times\sqrt N$ grid at
$\lambda/2$ spacing, $D_a(N)=(\sqrt N-1)\lambda/2$. The protection
region $\Gset\subset\Rset^3$ is compact: here a ground disk of
radius $L$ under an LEO formation at altitude $h_s$. The steering
vector is the \emph{exact spherical-wave (near-field) response of
the full array},
$[\ba_m(\bq)]_k=e^{-j(2\pi/\lambda)\|\bp_{m,k}-\bq\|}$,
$\ba(\bq)=[\ba_1^T,\ldots,\ba_M^T]^T\in\Cset^{\Mtot}$: no far-field
approximation across the array. The region
covariance and its $M^2$ blocks are
\begin{equation}\label{eq:Rg-def}
  \Rg=\frac{1}{|\Gset|}\int_\Gset \ba(\bq)\ba(\bq)^H d\bq,\quad
  \Rg^{(mn)}=\avg{\ba_m\ba_n^H},
\end{equation}
$\avg{\cdot}$ the average over $\Gset$. All $M^2$ blocks are
retained: $\Rg$ is \emph{not} block diagonal, the off-diagonal
blocks carrying the inter-platform near-field coupling. With
$\hat\bu_m(\bq)=(\bp_m-\bq)/\|\bp_m-\bq\|$ and $\bq_c$ the centroid
of $\Gset$, the angular radius of $\Gset$ from sub-array $m$ and its
maximum are
$\Delta\theta_{m}\triangleq\sup_{\bq\in\Gset}
\angle(\hat\bu_m(\bq),\hat\bu_m(\bq_c))$ and
$\Delta\theta_{\rm sep}\triangleq\max_m \Delta\theta_m$.

\begin{figure}[!tbp]
\centering
\begingroup
\resizebox{0.74\linewidth}{!}{%
\begin{tikzpicture}[font=\rmfamily, scale=1, transform shape, >=stealth,
  pan/.style={draw=black, thick, fill=gray!14},
  el/.style={fill=blue!55!black}]

\begin{scope}
\foreach \x in {-2.45,-1.75,-1.05,-0.35,0.35,1.05,1.75,2.45} {
  \pgfmathsetmacro{\y}{2.86-0.05*\x*\x}
  \draw[pan, rounded corners=0.5pt] (\x-0.115,\y-0.05) rectangle (\x+0.115,\y+0.05);
}
\draw[<->, thick] (-2.45,2.28) -- (2.45,2.28);
\node[font=\scriptsize, inner sep=1.2pt, anchor=north, fill=white, fill opacity=0.95, text opacity=1] at (0,2.23)
  {$M$ sub-arrays, span $D$};

\draw[orange!85!black, dashed, thick]
  ([shift={(70:3.28)}]0,0.10) arc[start angle=70, end angle=110, radius=3.28];
\node[orange!85!black, font=\tiny, align=left, anchor=east,
  fill=white, fill opacity=0.85, text opacity=1, inner sep=1.2pt]
  at (-1.58,3.00) {spherical across\\the formation};

\draw[blue!60!black, thick] (1.70,2.14) -- (2.02,1.98);
\draw[blue!60!black, thick] (1.82,2.36) -- (2.14,2.20);
\node[blue!60!black, font=\tiny, align=center, anchor=west,
  fill=white, fill opacity=0.88, text opacity=1, inner sep=1.1pt]
  at (2.30,2.20) {locally planar\\within one\\sub-array};

\draw[gray!55, thick] (-2.85,0) -- (2.95,0);
\fill[gray!30] (0,0) ellipse (0.82 and 0.15);
\draw[black, thick, dashed] (0,0) ellipse (0.82 and 0.15);
\draw[->, thin] (0,0) -- (0.78,0) node[midway, above, font=\tiny] {$L$};
\node[font=\scriptsize, anchor=north] at (-0.05,-0.20) {$\Gset$};

\coordinate (P) at (-1.05,2.75);
\fill[red!70!black, opacity=0.09] (P) -- (-0.82,0.0) -- (0.82,0.0) -- cycle;
\draw[red!70!black, dashed, thick] (P) -- (-0.82,0.0);
\draw[red!70!black, dashed, thick] (P) -- (0.82,0.0);
\draw[red!70!black, thick] ($(P)+(-80:0.60)$) arc[start angle=-80, end angle=-53, radius=0.60];
\node[red!70!black, font=\scriptsize] at ($(P)+(-69:1.37)$) {$\Delta\theta_m$};

\fill[red!80!black] (0,1.05) circle (2.2pt);
\node[font=\scriptsize, anchor=east, inner sep=2pt] at (-0.13,1.05) {$\bq^\star_{\rm in}$};
\fill[red!80!black] (2.55,0.92) circle (2.2pt);
\node[font=\scriptsize, anchor=south, inner sep=2pt] at (2.55,1.03) {$\bq^\star_{\rm sep}$};
\node[font=\scriptsize, align=center, anchor=north,
  fill=white, fill opacity=0.88, text opacity=1, inner sep=1.1pt]
  at (2.47,0.70) {angularly separated\\from $\Gset$};

\node[font=\small] at (0,-0.86) {(a)};
\end{scope}

\end{tikzpicture}%
}
\par\vspace{9pt}
\resizebox{0.48\linewidth}{!}{%
\begin{tikzpicture}[font=\rmfamily, scale=1, transform shape, >=stealth,
  pan/.style={draw=black, thick, fill=gray!14},
  el/.style={fill=blue!55!black}]

\begin{scope}
\fill[blue!45!black, opacity=0.08] (0,1.86) -- ++(230:2.20)
  arc[start angle=230, end angle=310, radius=2.20] -- cycle;
\fill[green!45!black, opacity=0.10] (0,1.86) -- ++(237.9:2.20)
  arc[start angle=237.9, end angle=302.1, radius=2.20] -- cycle;
\fill[red!70!black, opacity=0.20] (0,1.86) -- ++(252.7:2.20)
  arc[start angle=252.7, end angle=287.3, radius=2.20] -- cycle;

\draw[blue!45!black, thick] (0,1.86) -- ++(230:2.20);
\draw[blue!45!black, thick] (0,1.86) -- ++(310:2.20);
\draw[green!45!black, thick, dashed] (0,1.86) -- ++(237.9:2.20);
\draw[green!45!black, thick, dashed] (0,1.86) -- ++(302.1:2.20);
\draw[black!60, densely dotted, thick] (0,1.86) -- ++(250:2.20);
\draw[black!60, densely dotted, thick] (0,1.86) -- ++(290:2.20);
\draw[red!70!black, thick] (0,1.86) -- ++(252.7:2.20);
\draw[red!70!black, thick] (0,1.86) -- ++(287.3:2.20);

\node[red!70!black, font=\tiny, anchor=north,
  fill=white, fill opacity=0.82, text opacity=1, inner sep=0.8pt]
  at (0,0.73) {$\Delta\theta_{\rm sep}$};
\node[black!70, font=\tiny, anchor=west,
  fill=white, fill opacity=0.82, text opacity=1, inner sep=0.8pt]
  at (0.94,0.08) {$\lambda/(2D_a)$};
\node[green!45!black, font=\tiny, anchor=east,
  fill=white, fill opacity=0.82, text opacity=1, inner sep=0.8pt]
  at (-1.11,0.10) {$\frac{c^\star_{\square}}{\pi}\frac{\lambda}{D_a}$};
\node[blue!45!black, font=\tiny, anchor=west,
  fill=white, fill opacity=0.82, text opacity=1, inner sep=0.8pt]
  at (1.28,0.94) {$\lambda/D_a$};

\draw[pan] (-0.62,1.86) rectangle (0.62,3.10);
\foreach \i in {0,...,4} \foreach \j in {0,...,4}
  {\fill[el] (-0.48+0.24*\i,2.00+0.24*\j) circle (0.034);}
\draw[<->, thick] (-0.48,3.24) -- (0.48,3.24);
\node[font=\scriptsize, anchor=south, inner sep=0.5pt] at (0,3.28) {$D_a$};
\node[font=\scriptsize, align=left, anchor=west,
  fill=white, fill opacity=0.88, text opacity=1, inner sep=0.7pt]
  at (0.73,2.54) {$\sqrt N\!\times\!\sqrt N$\\grid};
\node[font=\scriptsize, anchor=north,
  fill=white, fill opacity=0.88, text opacity=1, inner sep=0.6pt]
  at (0,-0.56) {$c=1.36<\pi/2<c^\star_{\square}=2.52$};
\node[font=\small] at (0,-1.27) {(b)};
\end{scope}
\end{tikzpicture}%
}
\caption{Geometry and per-sub-array coherence mechanism. (a)~The $M$
compact sub-arrays span $D$: propagation is spherical across the
formation but locally planar within each sub-array. The protection
region $\Gset$ subtends $\Delta\theta_m$ at sub-array $m$, with
$\Delta\theta_{\rm sep}=\max_m\Delta\theta_m$; $\bq^\star_{\rm in}$
is in-cone, whereas $\bq^\star_{\rm sep}$ is angularly separated.
(b)~Angular scales for the LEO anchor. Here
$c=\pi D_a\Delta\theta_{\rm sep}/\lambda=1.36<\pi/2<c^\star_{\square}=2.52$,
so each diagonal covariance block remains in the one-dominant-mode
regime; the full-array simulation gives $r_{\rm eff}(\Rg)=M$ despite
$D\gg D_a$. Angles are enlarged for visual clarity.}
\label{fig:system}
\endgroup
\end{figure}

\subsection{Effective Rank}

For Hermitian $\bA$ with $\lambda_1\ge\lambda_2\ge\cdots$ we use the
half-power effective rank
$\reff(\bA)\triangleq|\{i:\lambda_i\ge\lambda_1/2\}|$ (a
\emph{relative} count: it is not monotone under compression). The only spectral tool needed is Poincar\'e separation (Cauchy
interlacing)~\cite[Sec.~4.3]{ref:hornjohnson}:
$\lambda_{r+1}(\bA)\le\lambda_1(\bm Q^H\bA\bm Q)$ for a compression
onto a subspace of codimension $r$, so if that compression has top
eigenvalue below $\lambda_1(\bA)/2$ then $\reff(\bA)\le r$.

\section{Main Results}
\label{sec:main}

Fix $\bq_0=\bq_c$, let $\ba_m^{(0)}=\ba_m(\bq_0)$,
$\bU=\mathrm{blkdiag}(\ba_1^{(0)},\ldots,\ba_M^{(0)})
\in\Cset^{\Mtot\times M}$ (so $\bU^H\bU=N\bI_M$),
$\Pill=\bU\bU^H/N$, $\Pdark=\bI-\Pill$ (rank $M(N-1)$). Factor
each entry \emph{exactly} as
$[\ba_m(\bq)]_k=e^{j\phi_m(\bq)}[\ba_m^{(0)}]_k\,
e^{j\theta_{m,k}(\bq)}$, where
$\phi_m(\bq)=-(2\pi/\lambda)(r_m(\bq)-r_m(\bq_0))$,
$r_m(\bq)=\|\bp_m-\bq\|$, is the platform-level spherical phase and
$\theta_{m,k}(\bq)\in(-\pi,\pi]$ the intra-sub-array excursion. The
single geometric input is the excursion bound
\begin{equation}\label{eq:A}
\textbf{(A)}\quad
|\theta_{m,k}(\bq)|\le\etatot\triangleq
\eta+\frac{2\pi R_a^2}{\lambda r_{\min}},\ 
\eta\triangleq\frac{2\pi R_a\Delta\theta_{\rm sep}}{\lambda},
\end{equation}
for all $m,k$, $\bq\in\Gset$, from a first-order path-length
expansion with an exact Fresnel remainder (Appendix); that remainder
is $1.3\times10^{-4}$ at the anchor of Section~\ref{sec:numerics}.
Every statement below holds verbatim with $\eta$ replaced by
$\etatot$, and we write $\eta$ for $\etatot$; the Slepian parameter
$c$ of Proposition~\ref{prop:slepian} obeys $c\le\eta$, with
$c=\eta$ for a linear sub-array and $c=\eta/\sqrt2$ for a square
panel with a disk region. No smallness of $\eta$ is assumed unless
stated.

\begin{lemma}[Coherence]\label{lem:coh}
Let $c_m(\bq)\triangleq\ba_m^{(0)H}\ba_m(\bq)/N$ and
$\br_m(\bq)\triangleq\ba_m(\bq)-c_m(\bq)\ba_m^{(0)}$. Then
$\ba_m^{(0)H}\br_m(\bq)=0$ and
$\|\br_m(\bq)\|^2=N(1-|c_m(\bq)|^2)$. Under \textup{(A)}: if
$\eta<\pi/2$ then $|c_m(\bq)|\ge\cos\eta>0$ and
$\|\br_m(\bq)\|^2\le N\sin^2\eta$; for every $\eta\ge0$,
$\|\br_m(\bq)\|^2\le N\eta^2$.
\end{lemma}

The role of $\pi/2$ is \emph{coherence}: for $\eta<\pi/2$ no element
drifts a quarter cycle from the reference profile, so the $N$
phasors cannot cancel; at $\eta=\pi/2$ cancellation ($c_m=0$) first
becomes possible. Thus $2R_a\Delta\theta_{\rm sep}<\lambda/2$ is a
worst-case coherence guarantee, a conservative \emph{design rule},
not the boundary of the collapse; the sharp threshold is
Proposition~\ref{prop:slepian}.

\begin{theorem}[Block-rank-1 collapse]\label{thm:rank1}
Let $\bc(\bq)=[c_1(\bq),\ldots,c_M(\bq)]^T$ and
$\br(\bq)=[\br_1(\bq);\ldots;\br_M(\bq)]$. For \emph{every}
$\eta\ge0$,
\begin{equation}\label{eq:decomp}
\ba(\bq)=\bU\bc(\bq)+\br(\bq),\qquad \bU^H\br(\bq)=\bm0,
\end{equation}
so $\Rg=\bU\bC\bU^H+\bX+\bX^H+\Rd$ with $\bC=\avg{\bc\bc^H}$,
$\bX=\avg{\bU\bc\,\br^H}$, and
$\Rd=\avg{\br\br^H}=\Pdark\Rg\Pdark$. Cauchy interlacing gives the
exact bounds $\lambda_i(\Rg)\ge N\lambda_i(\bC)$ for $i\le M$ and
\begin{equation}\label{eq:interlace}
\lambda_{M+1}(\Rg)\ \le\ \lambda_1(\Rd)\ \le\ \tr\Rd
= MN(1-\bar\kappa),
\end{equation}
with $\bar\kappa=\tr\bC/M\in[0,1]$ the mean squared coherence.
Hence $\reff(\Rg)\le M$ whenever $\bar\kappa>2M/(2M+1)$. Under
\textup{(A)} with $\eta<\pi/2$, Lemma~\ref{lem:coh} gives
$\lambda_1(\Rg)\ge N\cos^2\eta$ and
$\lambda_{M+1}(\Rg)\le MN\sin^2\eta$, so
\begin{equation}\label{eq:suff}
\tan^2\eta<\tfrac1{2M}\ \Longrightarrow\ \reff(\Rg)\le M
\ (M{=}8{:}\ \eta<0.245).
\end{equation}
\end{theorem}

Bound~\eqref{eq:interlace} holds for \emph{any} coupling $\bX$:
since $\Pdark\bU=\bm0$ the dark compression does not see $\bX$, so
inter-platform near-field coupling can never push $\lambda_{M+1}$
above $\lambda_1(\Rd)$. Condition~\eqref{eq:suff} is sufficient, not
necessary (Section~\ref{sec:numerics}: collapse persists to
$c\approx2.27$). \emph{Equality} $\reff=M$ additionally needs
$\lambda_M(\bC)\gtrsim\lambda_1(\bC)/2$, which holds at large
baselines where $\bC$ is diagonally dominant (Riemann--Lebesgue on
$e^{j(\phi_m-\phi_n)}$); in the coherent limit $\reff$ can drop
below $M$.

\begin{remark}[Distinct angular intervals per sub-array]
\label{rem:perblock}
With per-block references $\bq_{0,m}$, $\bU$ still obeys
$\bU^H\bU=N\bI_M$ and all proofs are unchanged, so each sub-array
may use its own local reference into $\Gset$, with a per-block
$\eta_m=2\pi R_a\Delta\theta_m/\lambda$: this covers sub-arrays
steered toward \emph{distinct} angular intervals of $\Gset$.
Sub-arrays protecting physically \emph{different} regions $\Gset_m$
define a different covariance, outside this model.
\end{remark}

\begin{proposition}[Exact per-block Slepian threshold]
\label{prop:slepian}
Let a sub-array be linear and in its far field with respect to
$\Gset$, with the direction-cosine offset of $\bq\in\Gset$ uniform
on $[-\Delta,\Delta]$, $\Delta\le\Delta\theta_{\rm sep}$. Then the
diagonal block is unitarily equivalent to the sinc-kernel (discrete
prolate, DPSS) matrix~\cite{ref:slepian} with
space--bandwidth parameter $c=\pi D_a\Delta/\lambda\le\eta$
\emph{exactly}, with equality when the modeled interval spans
$[-\Delta\theta_{\rm sep},\Delta\theta_{\rm sep}]$, and eigenvalues
$\frac{N\pi}{2c}\lambda_k(c)(1+o_N(1))$, $\lambda_k(c)$ the
Slepian--Pollak eigenvalues. Consequently:
\begin{enumerate}[label=(\roman*),leftmargin=2.25em,itemsep=0.4em,topsep=0.5em]
\item The per-sub-array
Shannon number (the classical count
$(2D_a/\lambda)\sin\theta$ of this
aperture) is $\sum_k\lambda_k(c)=2c/\pi\le2\eta/\pi$, equal to one
at $c=\pi/2$: the design rule
$D_a\Delta\theta_{\rm sep}<\lambda/2$ means \emph{fewer than one
spatial DoF per sub-array}.
\item The half-power block rank is one
iff $\lambda_1(c)/\lambda_0(c)<1/2$, i.e., iff
$c<c^\star=2.2772$ (finite $N$: $c^\star{=}1.99$ and $2.19$ for
$N_x{=}8$ and $27$ elements per side).
\item For a
square panel of side $D_a$ and a disk region the jinc kernel of the
disk band restricted to that \emph{square}~\cite{ref:slepian1964}
replaces the sinc, with $c=\pi D_a\Delta/\lambda=\eta/\sqrt2$,
Shannon number $c^2/\pi$ and threshold
$c^\star_{\square}=2.5227$ ($2.21$, $2.43$ for $N_x{=}8,27$), its
first excited level an exactly degenerate \emph{pair}.
\item For weakly coupled blocks (large baseline) the spectrum of
$\Rg$ is the union of the block spectra, so with a common $c$,
$\reff(\Rg)=M$ iff $c<c^\star$; unequal $c_m$ spread the transition
over the per-block thresholds.
\end{enumerate}
For non-negligible coupling, Theorem~\ref{thm:rank1}
instead supplies a coupling-independent upper bound through the
dark-subspace compression; no monotonicity of $\reff$ in the
coupling is claimed or needed.
\end{proposition}

\begin{corollary}[Beamforming ceiling and finite-$N$ law]
\label{cor:invq}
Let $\widetilde\bR=\Rg+\delta\bI$,
$\delta=\epsilon\lambda_1(\Rg)$, $\epsilon\ll1$ (loading /
finite-sample floor),
$\rhoeps=\ba_t^H\widetilde\bR^{-1}\ba_t$ ($\rhoeps\to\rhoavg$ as
$\delta\to0$), and $\bm\Pi_M$ the
top-$M$ eigenprojector of $\Rg$. Then
$\rhoeps=\rho_{\rm ill}+\rho_{\rm dark}$ with
$\rho_{\rm dark}\le\|\bm\Pi_M^{\perp}\ba_t\|^2/\delta$, and the dark
projection is
\begin{equation}\label{eq:darkproj}
\|\Pdark\ba_t\|^2=N\textstyle\sum_{m}(1-|g_m|^2),\quad
g_m\triangleq\tfrac1N\ba_m^{(0)H}\ba_{t,m},
\end{equation}
with $g_m$ the target--reference coherence at sub-array $m$
($|g_m|=1$ at $N=1$; the aperture pattern at the offset
$\Delta\bu_m=\hat\bu_m(\bq_t)-\hat\bu_m(\bq_0)$ for large $N$).
\begin{enumerate}[label=(\roman*),leftmargin=2.25em,itemsep=0.6em,topsep=0.5em]
\item \emph{Separated target}: in the weak-coupling regime
$\bm\Pi_M\approx\Pill$ ($\bC\approx\bI$, large baseline), if
$D_a|\Delta\bu_m|/\lambda\gtrsim1$ for all $m$ (every sub-array
resolves the target from $\Gset$), then
$|g_m|^2\ll1$, $\|\Pdark\ba_t\|^2=\Mtot(1-o(1))$, and once the dark
projection lies below the loading floor,
\begin{equation}\label{eq:MoverEps}
\rhoeps\approx
\frac{\sum_m|g_m|^2+\epsilon^{-1}\sum_m(1-|g_m|^2)}{\lambda_1(\bC)}
\longrightarrow \frac{M}{\epsilon\lambda_1(\bC)},
\end{equation}
i.e., $\rhoeps^{\rm dB}\approx10\log_{10}M+10\log_{10}
\epsilon^{-1}$ when $\lambda_1(\bC)\approx1$; $\lambda_1(\bC)<1$
raises the ceiling.
\item \emph{In-cone target}: if $|\Delta\bu_m|\ll\lambda/D_a$ for
all $m$ (range-only separation), then $|g_m|\approx1$ and, for
targets whose modal coefficients follow the concentration envelope
$|b_k|^2=O(\lambda_k)$, block mode $k$ carries dark energy
$O(\lambda_k(c))$ and contributes $O(1)$ regardless of $\epsilon$:
$\rhoeps=O(M K_\epsilon)$,
$K_\epsilon=\#\{k:\lambda_k\ge\epsilon\lambda_0\}=O(\log
\epsilon^{-1})$~\cite{ref:landauWidom1980} (no $1/\epsilon$
amplification, for any $N$).
\end{enumerate}
\end{corollary}

Corollary~\ref{cor:invq} separates effective rank from finite-$N$
behavior: the rise of $\rhoeps$ with $N$ is not a rise of $\reff$
(which stays $M$) but the growth of the dark projection from $0$ at
$N=1$ toward its aperture limit, and its migration below the loading
floor as $D_a(N)$ resolves the target.

\begin{proposition}[Worst-case leakage and duality]\label{prop:dual}
Let $\Pset(\Gset)$ be the probability measures on $\Gset$,
$\bR_\mu=\int\ba\ba^H d\mu$, and
${\rm SIR}_{\rm wc}(\bw)=|\ba_t^H\bw|^2/\max_{\bq\in\Gset}
|\ba(\bq)^H\bw|^2$. For every $\bw$ and every $\mu\in\Pset(\Gset)$
with $\bw^H\bR_\mu\bw>0$, a supremum dominates a mean, so
\begin{equation}\label{eq:wcbound}
{\rm SIR}_{\rm wc}(\bw)\le
\frac{|\ba_t^H\bw|^2}{\bw^H\bR_\mu\bw},
\ \text{hence}\ \ 
{\rm SIR}_{\rm wc}(\bw)\le\rhoavg
\end{equation}
the second at uniform $\mu$, attained by $\bw\propto\Rg^{-1}\ba_t$
when $\Rg\succ0$. No inverse of a general $\bR_\mu$ appears:
$\Pset(\Gset)$ contains Diracs, for which $\bR_\mu$ is singular.
So $\rhoavg$ is the exact average-leakage optimum and a one-sided
upper bound on worst-case SIR, not a value we claim is attained.
Since~\eqref{eq:decomp} is pointwise in $\bq$, the same structural
rank restriction applies to $\bR_\mu$ under any probability
weighting of $\Gset$.
\end{proposition}

\begin{corollary}[Baseline and architecture]\label{cor:agnostic}
\label{cor:baseline}
\begin{enumerate}[label=(\roman*),leftmargin=2.25em,itemsep=0.6em,topsep=0.5em]
\item The bounds of Theorem~\ref{thm:rank1} and the thresholds of
Proposition~\ref{prop:slepian} depend on
$(D_a,\Delta\theta_{\rm sep},\lambda)$, not on $D$: the sufficient
collapse condition, and hence $\reff\le M$ wherever it holds, is
baseline-independent. The ceiling~\eqref{eq:MoverEps} is
baseline-independent in the separated-target regime ($|g_m|\ll1$);
otherwise $D$ enters $\rhoeps$ only through $g_m(D)$.
\item $\reff\le M$ is a property of $\Rg$ (and of every $\bR_\mu$),
not of the beamformer: MVDR, MMSE, leakage-constrained MRT and the
worst-case optimum of~\eqref{eq:socp} (any
$\bw=f(\bR_\mu,\ba_t)$) face the same $M$-dimensional illuminated
subspace and the same ceilings.
\end{enumerate}
\end{corollary}

\section{Numerical Validation}
\label{sec:numerics}

\subsection{Simulation Setup}
$M=8$ satellites at $h_s=600$~km,
$x_m\in\{\pm21.4,\pm64.3,\pm107.1,\pm150\}$~km ($D=300$~km),
$\lambda=0.15$~m ($2$~GHz); $\Gset$: ground disk of radius
$L=20$~km, $2\times10^4$ equal-area points (doubling changes the
$N{=}64$ results by $<0.03$~dB); sub-arrays: $\sqrt N\times\sqrt N$
grids at $\lambda/2$, $N\in\{1,4,16,64,256,729\}$, $D_a$ up to
$1.95$~m; loading $\epsilon=10^{-6}$. Targets at $10$~km altitude:
$\bq^\star_{\rm sep}=(300,0,10)$~km, with
$D_a|\Delta\bu_m|/\lambda\in[5.3,6.4]$ at $N=729$
(Corollary~\ref{cor:invq}(i)), and $\bq^\star_{\rm in}=(0,0,10)$~km
above the disk center, inside every cone (regime~(ii)).
$\Delta\theta_{\rm sep}=0.0333$~rad ($1.9^\circ$) gives, at
$D_a=1.95$~m ($R_a=1.38$~m), $\eta=1.92$ and $c=1.36$: \emph{above}
both $\eta<0.245$ (\eqref{eq:suff}, $M=8$) and $\eta<\pi/2$, well
\emph{below} $c^\star_{\square}(27)=2.43$, so $\reff=M$ here is an
empirical fact in the regime of Proposition~\ref{prop:slepian}, not
a consequence of~\eqref{eq:suff}.

\subsection{Rank Collapse and the Capon Ceiling}

$\reff(\Rg)=8=M$ for every $N$; the ninth eigenvalue stays at
$0.0007$--$0.17$ of $\lambda_1$ from $N=4$ to $729$. For $\bq^\star_{\rm sep}$, $\rhoeps$ climbs
$9.1\to37.2\to62.4\to68.3\to69.1\to70.1$~dB ($N=1\to729$),
following~\eqref{eq:MoverEps} within $0.7$~dB for $N\ge64$, where the
separation hypothesis first holds ($D_a|\Delta\bu_m|/\lambda=1.4$--$1.7$,
against $0.2$--$0.7$ at $N\le16$, so~\eqref{eq:MoverEps} does not
apply there). The ceiling is $M/(\epsilon\lambda_1(\bC))$: $\lambda_1(\bC)$ falls from
$1.02$ at $N=1$ to $0.76$ at $N=729$ as each sub-array pattern rolls
off across $\Gset$, lifting the ceiling from $68.9$ to $70.2$~dB,
which is why $\rhoeps$ passes $10\log_{10}(M/\epsilon)=69.0$~dB. For $\bq^\star_{\rm in}$ it
stays at $9.0$--$16.8$~dB even at $\Mtot=5832$: with
$|g_m|\ge0.9955$ no $1/\epsilon$ amplification occurs, as
Corollary~\ref{cor:invq}(ii) states.

\begin{figure}[H]
\centering
\includegraphics[width=0.92\linewidth]{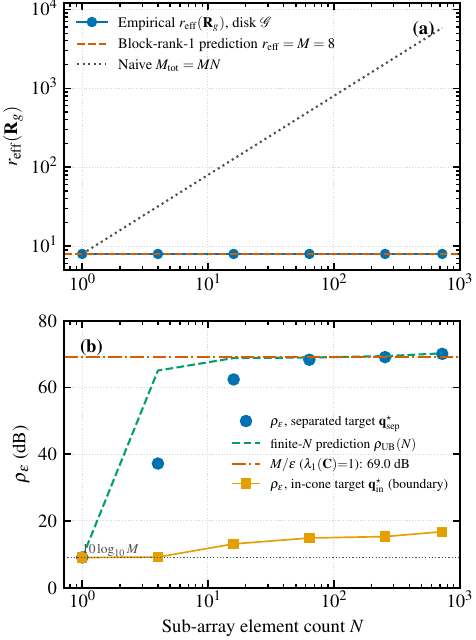}
\caption{Disk region, $M{=}8$, $\epsilon{=}10^{-6}$.
(a)~$\reff(\Rg)$ against the naive count $MN$. (b)~$\rhoeps$
vs.\ $N$: $\bq^\star_{\rm sep}$ (dots) with prediction~\eqref{eq:MoverEps}
(dashed) and $\bq^\star_{\rm in}$ (squares); dash--dot is
$10\log_{10}(M/\epsilon)$, the ceiling at $\lambda_1(\bC)=1$.}
\label{fig:reff_rho}
\end{figure}

\subsection{Beamformers and Worst-Case SIR}

MVDR, MMSE ($\sigma_n^2=\lambda_{\min}/100$) and SLNR realize
identical suppression (within $10^{-4}$~dB at $\bq^\star_{\rm sep}$)
per Corollary~\ref{cor:agnostic}(ii). The \emph{worst-case} SIR of
loaded MVDR, from a two-pass refined scan of $\Gset$ (refinement
$\le0.25$~dB; analytic envelope within $1.3$~dB), is
$22.5/48.2/66.8/69.5/73.4$~dB at $N=4/16/64/256/729$ for
$\bq^\star_{\rm sep}$, $9$--$21$~dB below the average SIR
${\rm SIR}_{\rm avg}(\bw_\epsilon)$ of the same weight
($37.2$--$94.1$~dB), as~\eqref{eq:wcbound} requires. Four quantities
differ: Proposition~\ref{prop:dual} bounds
${\rm SIR}_{\rm wc}(\bw_\epsilon)$ by the \emph{unloaded}
$\rhoavg$ (${\ge}94$~dB here), not by
$\rhoeps$, whose ceiling is not a worst-case bound, so
${\rm SIR}_{\rm wc}=73.4$~dB may exceed $\rhoeps=70.1$~dB.
Suppression at $\bq^\star_{\rm sep}$ is therefore real in the
worst-case sense; at $\bq^\star_{\rm in}$ the worst case never
exceeds $1$~dB at any $N$, so inside the cone $\rhoeps$ must be read
as an average-leakage metric.

\begin{figure}[H]
\centering
\includegraphics[width=0.90\linewidth]{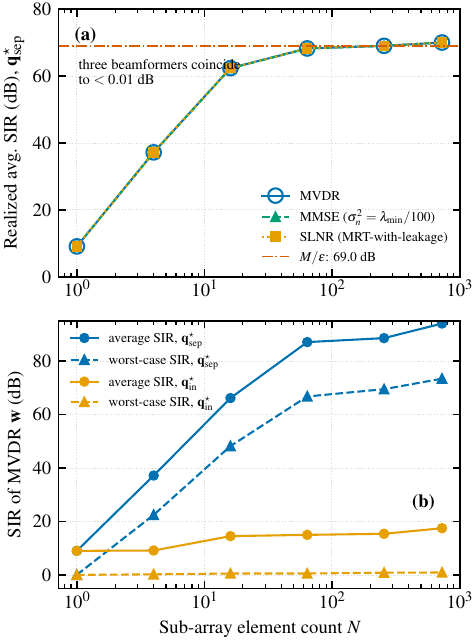}
\caption{(a)~Average SIR at $\bq^\star_{\rm sep}$ for MVDR, MMSE
and SLNR. (b)~${\rm SIR}_{\rm avg}$ and ${\rm SIR}_{\rm wc}$ of one
loaded MVDR weight over $\Gset$: realized SIRs, not $\rhoeps$.}
\label{fig:beamformers}
\end{figure}

\subsection{Concentration Threshold}
 Sweeping
$f=0.3$--$60$~GHz at fixed $D_a=1.95$~m, $N=64$ traverses
$c=0.20$--$40.8$. $\reff=M$ through $c=2.25$ and departs at
$c=2.28$; the ninth-eigenvalue half-power crossing is $c=2.27$,
$2.7\%$ above $c^\star_{\square}(8)=2.21$, the gap being the $3\%$
spread of the per-platform angular radii about $L/h_s$, which also
staggers the per-block transitions so that $\reff$ reaches $3M$ only
at $c=2.48$. Element spacing exceeds $\lambda/2$ for
$f\gtrsim0.54$~GHz, so per-block grating lobes are present
throughout, yet the block operator depends only on $(c,N_x)$.

\begin{figure}[H]
\centering
\includegraphics[width=0.86\linewidth]{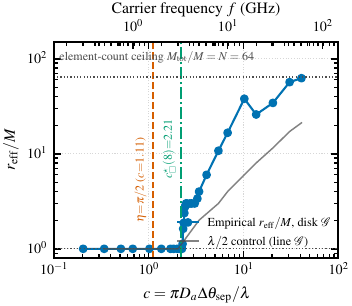}
\caption{$\reff/M$ vs.\ the block concentration
$c=\pi D_a\Delta\theta_{\rm sep}/\lambda$
(disk region; fixed $D_a=1.95$~m, $N=64$, $f=0.3$--$60$~GHz).
Dash--dot: the $8\times8$ threshold $c^\star_{\square}$
(Proposition~\ref{prop:slepian}(iii)); dashed: the rule
$\eta=\pi/2$, i.e.\ $c=\pi/(2\sqrt2)=1.11$ on this axis.
Gray: $\lambda/2$ control, line region.}
\label{fig:phase-trans}
\end{figure}

\subsection{Baseline Sweep and Other Applications}
At $N=64$, sweeping $D=50$--$3000$~km leaves
$\reff(\Rg)=8$ throughout while the hull estimate
$\rprolate=D^2\Omega_\Gset/\lambda^2$ grows from $3.9\times10^{8}$
to $1.4\times10^{12}$: the ambient prolate dimension never binds.
$\rhoeps(\bq^\star_{\rm sep})=68.5/68.5/68.3/66.9/62.8$~dB at
$D=50/100/300/1000/3000$~km, near-invariant to $300$~km and then
decreasing, not through $\reff$ but because outer platforms move
toward the target's zenith (offset $1.6\to0.22$ beamwidths): the
\emph{separation hypothesis} of Corollary~\ref{cor:invq}(i)
degrades, as Corollary~\ref{cor:baseline}(i) anticipates. The same
diagnostic places an ISAC roadside cluster ($1.0$~m, $6$~GHz,
$0.5^\circ$) at $c=0.55$ and a cell-free cluster ($0.3$~m,
$10$~GHz, $1.5^\circ$) at $c=0.82$, far below $c^\star_{\square}$.

\section{Discussion and Limitations}
\label{sec:discussion}

\subsection{Grating Lobes}
Within each $\lambda/2$-spaced sub-array there
are none; at the platform level the $\gg\lambda$ baselines produce
an angular grating comb set by the platform \emph{spacing}
$d\approx D/(M-1)$, not by $D$: ground pitch $\lambda h_s/d$,
$0.2$--$12.6$~m here, the aliasing price of a sparse
aperture~\cite{ref:bjornsonReality2019}; modular arrays keep such
\emph{angular} lobes even in near-field
focusing~\cite{ref:liModularFocus2024}. Range ambiguity is weaker
but not excluded: a filled aperture's quadratic element phase does
not re-cohere at a second range~\cite{ref:hussainThinning2026},
while a sparse formation samples that profile at $M$ points only, so
range--angle ambiguity must be checked per geometry. The collapse is indifferent
to the comb ($\reff=8$ while the pitch varies $60\times$): grating
lobes redistribute \emph{where} leakage falls; they do not enlarge
the illuminated subspace.

\subsection{Hybrid Arrays}
The results act on the spectrum of $\Rg$, not
on how $\bw$ is realized. A sub-connected hybrid base station (one RF chain per
compact $\lambda/2$ panel~\cite{ref:liModularFocus2024}) is
structurally the $M\times N$ model here, with only $M$ digital
variables. For a hybrid architecture the operative covariance is
$\bm{F}_{\rm RF}^H\Rg\bm{F}_{\rm RF}$, whose algebraic rank cannot
exceed $M$ under exact collapse; its \emph{half-power} rank depends
also on the conditioning $\bm{F}_{\rm RF}$ introduces, since
compression can lower $\lambda_1$. The element-domain result is a DoF
upper envelope, not an attainability claim.

\subsection{Near-Field Boundary}
The analysis requires each sub-array in
its far field, $h_s\gg2(2R_a)^2/\lambda=101$~m, while the collective
near field holds at any practical distance
($2D^2/\lambda=1.2\times10^{9}$~km). Inside it the statements stay
range-dependent through
$c\propto\Delta\theta_{\rm sep}\approx L/h_s$: \emph{descending}
breaks the collapse first ($c>c^\star_{\square}$), ascending deepens
it. Exact spherical steering exists at every finite range, so
focusing never simply ceases; what degrades is the useful
\emph{range discrimination}, limited by beam depth and the effective
beamfocusing Rayleigh
distance~\cite{ref:abdallahPolar2025}. It resides
in the $M$ platform-level phases ($\bC$), and it and the separation
hypothesis of Corollary~\ref{cor:invq}(i) must be rechecked at each
range.

\subsection{Scope and Limitations}
$\Gset$ is assumed known; the treatment is
single-snapshot and narrowband. With finite snapshots
$\reff\approx M\ll\Mtot$ lets the Reed--Mallett--Brennan rule need
only $O(M)$ samples for the illuminated subspace, while dark-subspace
error feeds $\epsilon$. Mutual coupling acts per sub-array
as a fixed map common to all $\bq$, preserving block-rank-1.

\section{Conclusion}
\label{sec:conclusion}

For sparse arrays of $M$ coherent sub-arrays, per-sub-array
coherence can restrict the region covariance to an $M$-dimensional
illuminated subspace: interlacing gives the nonasymptotic sufficient
condition $\tan^2\eta<1/(2M)$, and concentration theory the less
conservative per-block thresholds $c^\star$ and
$c^\star_{\square}$. In that regime the regularized Capon functional
approaches $M/(\epsilon\lambda_1(\bC))$ for angularly resolved
targets and shows no $1/\epsilon$ amplification for in-cone ones.
That scaling is covariance-side and beamformer-independent, though no
particular front end need attain it. The LEO study confirms both laws, including
realized worst-case SIR.

\appendix
\section{Proofs of the Main Results}
\label{app:proof}

\subsection{Geometric Excursion Bound (A)}
With
$\|\bp_{m,k}-\bq\|=r_m(\bq)+\bm{\xi}_k^T\hat\bu_m(\bq)
+\varrho_{m,k}$, $|\varrho_{m,k}|\le\|\bm{\xi}_k\|^2/(2r_m)$,
Cauchy--Schwarz with
$\|\hat\bu_m(\bq)-\hat\bu_m(\bq_0)\|\le\Delta\theta_{\rm sep}$
gives
$|\theta_{m,k}|\le2\pi R_a\Delta\theta_{\rm sep}/\lambda
+2\pi R_a^2/(\lambda r_{\min})$.

\subsection{Proof of Lemma~\ref{lem:coh}}
$c_m=e^{j\phi_m}\tilde c_m$,
$\tilde c_m=\frac1N\sum_ke^{j\theta_{m,k}}$; orthogonality is by
construction, $\|\br_m\|^2=N(1-|c_m|^2)$ is Pythagoras. If
$\eta<\pi/2$: $|\tilde c_m|\ge\operatorname{Re}\tilde c_m=
\frac1N\sum_k\cos\theta_{m,k}\ge\cos\eta>0$, so
$1-|c_m|^2\le\sin^2\eta$. For all $\eta$, $\cos x\ge1-x^2/2$ gives
$|\tilde c_m|\ge1-\eta^2/2$; if $\eta<1$ that bound is positive and
squaring it gives $1-|c_m|^2\le\eta^2-\eta^4/4\le\eta^2$, while for
$\eta\ge1$ the claim is immediate from
$1-|c_m|^2\le1\le\eta^2$.\hfill$\square$

\subsection{Proof of Theorem~\ref{thm:rank1}}
Stacking
$\ba_m=c_m\ba_m^{(0)}+\br_m$, $\ba_m^{(0)H}\br_m=0$
gives~\eqref{eq:decomp}; expanding
$\avg{(\bU\bc+\br)(\bU\bc+\br)^H}$ with $\Pdark\bU=\bm0$,
$\Pdark\br=\br$ yields the decomposition and
$\Pdark\Rg\Pdark=\Rd$. Compressing by $\bm Q=\bU/\sqrt N$:
$\bm Q^H\Rg\bm Q=N\bC$ (terms with $\br$ vanish), so
$\lambda_i(\Rg)\ge N\lambda_i(\bC)$, $i\le M$, by Poincar\'e
separation~\cite[Sec.~4.3]{ref:hornjohnson}; compressing onto
$\Range(\Pdark)$ (codimension $M$) gives
$\lambda_{M+1}(\Rg)\le\lambda_1(\Rd)\le\tr\Rd$. Pointwise Pythagoras gives $\tr\Rd=MN(1-\bar\kappa)$; with
$\lambda_1(\bC)\ge\bar\kappa$, $\lambda_{M+1}<\lambda_1/2$ once
$\bar\kappa>2M/(2M+1)$.
Lemma~\ref{lem:coh} gives $\bar\kappa\ge\cos^2\eta$, and
$MN\sin^2\eta<N\cos^2\eta/2\iff\tan^2\eta<1/(2M)$.\hfill$\square$

\subsection{Proof of Proposition~\ref{prop:slepian}}
In the sub-array far field
the intra-sub-array excursion is linear in the element abscissa,
$\theta_{m,k}(\bq)=-(2\pi/\lambda)x_k\sigma(\bq)$, with $\sigma$ the
direction-cosine offset of $\bq$ along the array axis,
$|\sigma|\le\Delta$. Averaging $\ba_m\ba_m^H$ over $\sigma$ uniform
on $[-\Delta,\Delta]$ gives, entrywise,
$[\Rg^{(mm)}]_{kl}=[\ba_m^{(0)}]_k\overline{[\ba_m^{(0)}]_l}\,
\frac{1}{2\Delta}\int_{-\Delta}^{\Delta}
e^{-j\frac{2\pi}{\lambda}(x_k-x_l)\sigma}d\sigma$, i.e.,
$\Rg^{(mm)}=\bm D\bm S\bm D^H$ with $\bm D=\mathrm{diag}(\ba_m^{(0)})$
unitary, $t_k=2x_k/D_a\in[-1,1]$,
$S_{kl}=\operatorname{sinc}(c(t_k-t_l))$ and $c=\pi D_a\Delta/\lambda$:
the DPSS matrix~\cite{ref:slepian}. For $\{t_k\}$ uniform with
weight $2/N$, $\bm S$ is the Nystr\"om discretization of
$(\mathcal{K}f)(t)=\int_{-1}^{1}\operatorname{sinc}(c(t-s))f(s)ds
=\frac{\pi}{c}(\mathcal{Q}_cf)(t)$, $\mathcal{Q}_c$ the prolate
operator of kernel $\frac{\sin c(t-s)}{\pi(t-s)}$, whence
$\lambda_k(\bm S)=\frac{N\pi}{2c}\lambda_k(c)(1+o_N(1))$.
\par\medskip
oindent	extbf{(i) Shannon number.} $\mathcal{Q}_c$ has kernel diagonal $c/\pi$, so
$\sum_k\lambda_k(c)=\tr\mathcal{Q}_c=2c/\pi$; equivalently, the
phase-space volume $(2)(2c)/(2\pi)$~\cite{ref:landauWidom1980}.
Writing $c=\pi D_a\Delta/\lambda$ recovers the classical aperture
count $2D_a\Delta/\lambda$, and $\sum_k\lambda_k(c)=1$ exactly at
$c=\pi/2$. \par\medskip
oindent	extbf{(ii) Half-power block threshold.} $\lambda_1(c)/\lambda_0(c)$ increases from $0$
($c\to0$) to $1$ ($c\to\infty$), so $\lambda_1<\lambda_0/2$ holds on
a maximal interval $c<c^\star$; solving
$\lambda_1(c)=\lambda_0(c)/2$ on the classical spectrum gives
$c^\star=2.2772$, and the same equation on $\lambda_k(\bm S)$ gives
the finite-$N$ values quoted. \par\medskip
oindent	extbf{(iii) Square panel and disk.} For a square panel of side
$D_a$ and a disk region, averaging over a disk of radius $\Delta$ in
direction cosines replaces the sinc by the jinc kernel
$2J_1(c\|\bm t_k-\bm t_l\|)/(c\|\bm t_k-\bm t_l\|)$,
$\bm t_k=2\bm\xi_k/D_a\in[-1,1]^2$, $c=\pi D_a\Delta/\lambda$:
the disk band restricted to the \emph{square}, not Slepian's
rotationally symmetric disk operator~\cite{ref:slepian1964}. Its
phase-space volume is $(4)(\pi c^2)/(2\pi)^2=c^2/\pi$; the panel's
symmetry group has a two-dimensional irreducible representation
carrying both first-excited modes, so they are exactly degenerate;
and $\lambda_1(c)=\lambda_0(c)/2$ gives
$c^\star_{\square}=2.5227$, or $2.21$ and $2.43$ on the
$N_x\times N_x$ jinc matrix for $N_x{=}8,27$. Since
$\|\bm\xi_k\|\le R_a=D_a/\sqrt2$, $c=\eta/\sqrt2$. \par\medskip
oindent	extbf{(iv) Weak-coupling limit.} For $m
e n$,
$[\Rg^{(mn)}]_{kl}=\avg{e^{j(\phi_m-\phi_n)}
e^{j(\theta_{m,k}-\theta_{n,l})}}$; at large baseline
$\phi_m-\phi_n=-(2\pi/\lambda)(r_m-r_n)$ sweeps many cycles across
$\Gset$ while the second factor varies slowly, so the average
vanishes by Riemann--Lebesgue and
$\Rg\to\mathrm{blkdiag}(\Rg^{(11)},\ldots,\Rg^{(MM)})$ in norm. By
Weyl's inequality the spectrum converges to the union of the $M$
identical block spectra, each contributing one eigenvalue above
half power exactly when $c<c^\star$; hence $\reff(\Rg)=M$ iff
$c<c^\star$.\hfill$\square$

\subsection{Proof of Proposition~\ref{prop:dual}}
$\bw^H\bR_\mu\bw=\int|\ba^H\bw|^2d\mu\le
\max_\Gset|\ba^H\bw|^2$ ($\mu$ a probability measure) gives the
first bound in~\eqref{eq:wcbound}; the second is its uniform-$\mu$
case followed by the supremum defining $\rhoavg$, which needs no
inverse. If $\Rg\succ0$, Cauchy--Schwarz in the $\Rg$ inner product
evaluates that supremum as $\ba_t^H\Rg^{-1}\ba_t$, tight iff
$\bw\propto\Rg^{-1}\ba_t$ (Capon). The last claim holds
because~\eqref{eq:decomp} and Lemma~\ref{lem:coh} are pointwise in
$\bq$ and integrate against any $\mu$.\hfill$\square$

\subsection{Proof of Corollary~\ref{cor:invq}}
The split is the spectral theorem
for $\widetilde\bR^{-1}$; $\Pdark$ is block diagonal, so
$\|\Pdark\ba_t\|^2=\sum_m(N-|\ba_m^{(0)H}\ba_{t,m}|^2/N)$,
which is~\eqref{eq:darkproj}; $\bm\Pi_M$ coincides with the range
of $\bU$ only when $\bX=\bm0$, the two projections differing by at
most $2\|\ba_t\|^2\|\bm\Pi_M-\bU\bU^H\!/N\|_2$,
so~\eqref{eq:MoverEps} is stated for $\bC\approx\bI$ (large
baseline). \par\medskip
oindent	extbf{(i) Separated target.} $\delta\approx\epsilon N\lambda_1(\bC)$ yields~\eqref{eq:MoverEps}
once the dark projection lies below the floor. \par\medskip
oindent	extbf{(ii) In-cone target.} For a target
whose modal coefficients follow the concentration envelope,
$|b_k|^2=O(\lambda_k)$~\cite{ref:slepian}, mode $k$ contributes
$O(\lambda_k/(\lambda_k+\epsilon\lambda_0))=O(1)$.\hfill$\square$

\bibliographystyle{IEEEtran}
\bibliography{wcl_refs_v4}

\end{document}